\documentclass[a4paper,11pt]{article}
\usepackage{float}
\usepackage{xcolor}
\usepackage{jheppub}
\usepackage{multirow}

\newcommand{\jpsi}{\ensuremath{J/\psi}}
\newcommand{\psip}{\ensuremath{\psi(2S)}}
\newcommand{\bbar}{\ensuremath{\overline{\bf B}}}
\newcommand{\bb}{\ensuremath{{\bf B}}}
\newcommand{\calA}{\ensuremath{{\cal A}}}

\newcommand{\SUF}{\ensuremath{\mathrm{SU}(3)_{F}}}
\newcommand{\KSD}{\ensuremath{k_{{\rm SD}}}}

\title{\boldmath\texorpdfstring{ Negative-$\Sigma\bar\Sigma$ Angular-Parameter Puzzle in $\jpsi$ and $\psip$  Decays}{The Negative Sigma-Sigma-bar Angular-Parameter Puzzle in $\jpsi$ and $\psip$  Decays}}

\author{Chao-Qiang Geng}
\author{Xiang-Nan Jin}
\author{Chia-Wei Liu}

\affiliation{School of Fundamental Physics and Mathematical Sciences,
Hangzhou Institute for Advanced Study, UCAS, Hangzhou 310024, China}

\emailAdd{cqgeng@ucas.ac.cn}
\emailAdd{xnjin@ucas.ac.cn}
\emailAdd{chiaweiliu@ucas.ac.cn}

\abstract{Experimental measurements of $\jpsi$ and $\psip$ decays to octet baryon pairs
reveal a negative-$\Sigma\bar\Sigma$ puzzle in the angular distributions:
$\alpha_\Sigma^{\jpsi}<0$, whereas the other measured $\jpsi$ octet
channels and the corresponding $\psip$ channels have positive angular
parameters.  We show that an \SUF-exact singlet description is strongly
disfavored in a $\chi^2$ fit, giving $\chi^2\simeq10^5$.  By contrast, a
reduced scenario in which a common final-state interaction mixes the
${\rm S}$- and ${\rm D}$-wave amplitudes, with an antisymmetric direct
octet and a full diagonal ${\rm S}$--${\rm D}$ rescattering block, gives
$\chi^2=4.21$ for two nominal degrees of freedom.  The negative
$\Sigma\bar\Sigma$ angular parameter is traced to the pattern of the
rescattered ${\rm S}$- and ${\rm D}$-wave amplitudes.The apparent
large \SUF-breaking effect is decomposed into three ingredients: the fitted
${\rm D}$-wave source is much larger than the ${\rm S}$-wave source, the
breaking remains modest when normalized to the ${\rm D}$ wave, and
${\rm D}\to{\rm S}$ rescattering transfers the \SUF-breaking effect from the
${\rm D}$ wave into the small ${\rm S}$-wave baseline, where it appears
large.  The largest remaining pull comes from the
$\psip\to N\bar N$ angular input, motivating a renewed
experimental check of $\psip\to n\bar n$.}


\hypersetup{
	pdftitle={The Negative Sigma-Sigma-bar Angular-Parameter Puzzle in J/psi and psi(2S) Decays},
	pdfauthor={Chao-Qiang Geng, Xiang-Nan Jin, Chia-Wei Liu},
	pdfkeywords={Charmonium, flavor symmetry, final-state interactions, baryon-antibaryon production}
}

\begin{document}

	\maketitle
	\suppressfloats[t]

	\section{Introduction}

\begin{figure}[!htbp]
		\centering
		\begin{minipage}[c]{0.5\textwidth}
			\centering
			\includegraphics[width=\linewidth]{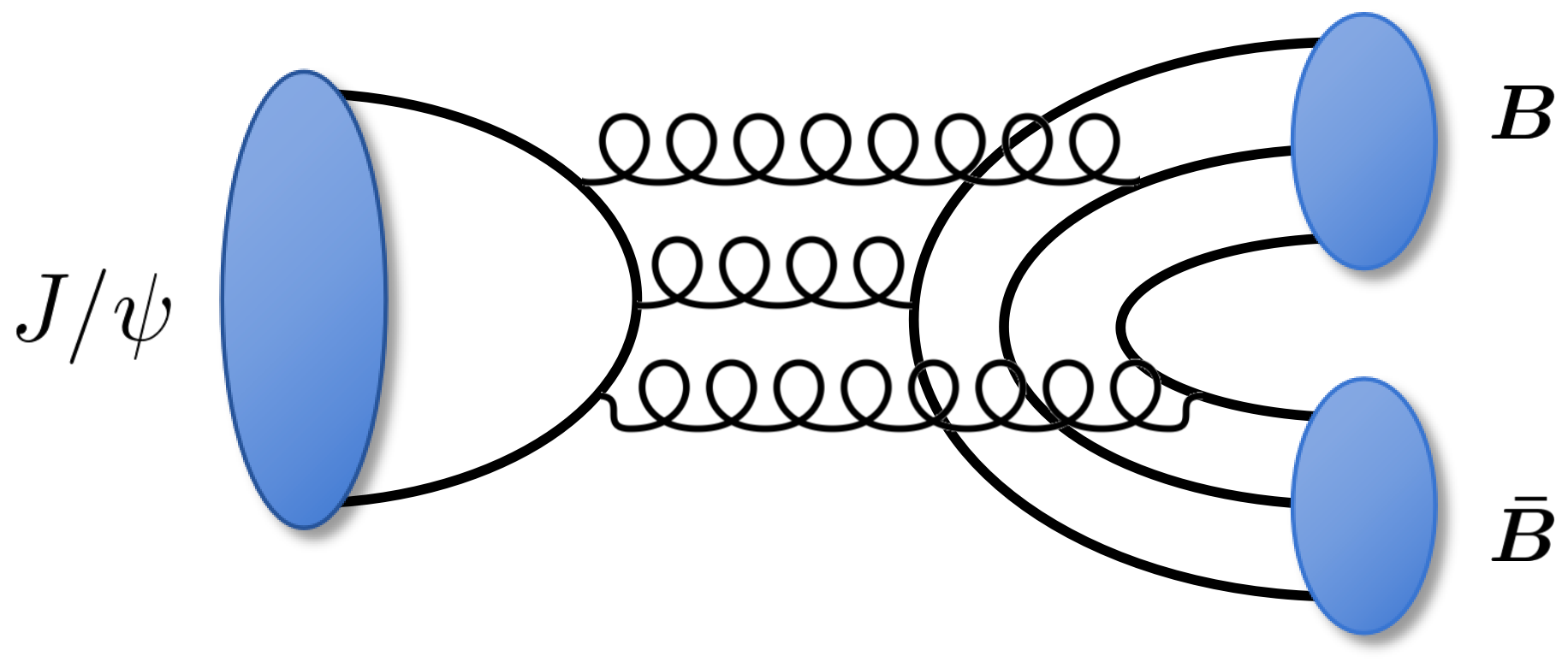}
		\end{minipage}
~~
		\begin{minipage}[c]{0.3 \textwidth}
			\centering
			\includegraphics[width=\linewidth]{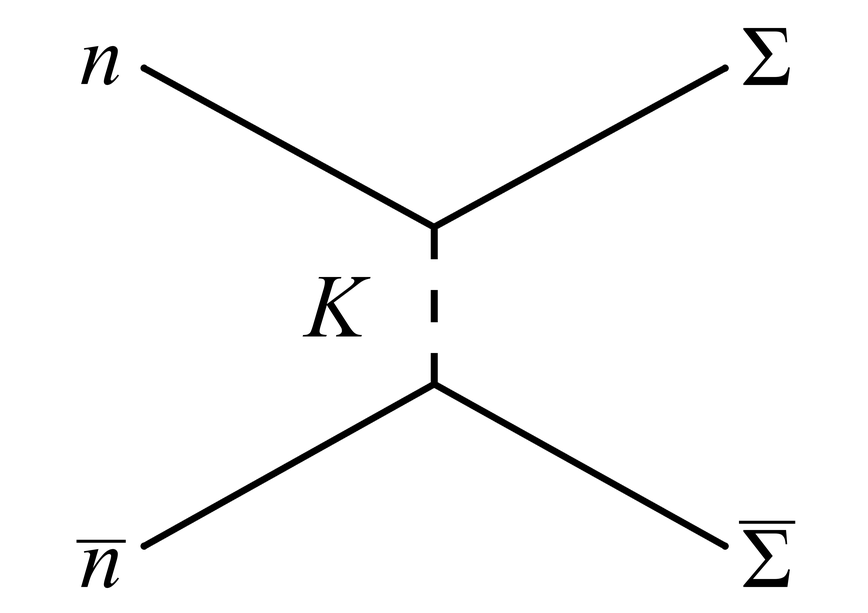}
		\end{minipage}
		\caption{Direct production and subsequent rescattering.  Left:
			three-gluon topology for $\jpsi,\psip\to\bb\bbar$. where $\bb$ denotes an octet baryon.  Right:
			$n\bar n\to\Sigma\bar\Sigma$ through $K$-exchange, representing
			the FSI transition following direct production.}
		\label{fig:production-fsi-schematic}
	\end{figure}

	The experimental angular distributions for baryonic $\jpsi$ and $\psip$ decays 
	show an anomalous sign pattern in the $\Sigma\bar\Sigma$ channel.  Here
	and below $N=p,n$, $\Sigma=\Sigma^{+,0,-}$, and
	$\Xi=\Xi^{0,-}$.  In $\jpsi$ decays the data find
	$\alpha_\Sigma^{\jpsi}<0$, whereas the measured $N\bar N$,
	$\Lambda\bar\Lambda$, and $\Xi\bar\Xi$ channels, as well as the baryonic
	decays of $\psip$, have positive
	angular parameters~\cite{BESIII:2017kqw,BESIII:2018cnd,
	BESIII:2020fqg,BESIII:2021ypr,BESIII:Sigma0PRL2024,
	BESIII:LambdaPRL2022,ParticleDataGroup:2024cfk,BESIII:Xi0JpsiPhase2023,BESIII:Xi0Psi2SPhase2026,
	BESIII:Xi1530Psi2S2026,BESIII:XiMinusJpsi2026,
	BESIII:CharmoniumSymmetry2026}.
	The relevant data are collected in Table~\ref{tab:experimental-inputs}.
	The phase measurements sharpen this observation.  In $\jpsi$
	decay, $\Delta\Phi_{\Sigma}$ is negative, whereas the measured
	$\Lambda\bar\Lambda$ and $\Xi\bar\Xi$ phases are positive.  Thus the
	$\Sigma\bar\Sigma$ channel is singled out both by the sign of
	$\alpha_\Sigma^{\jpsi}$ and by a distinct strong-phase pattern.  This makes
	final-state interaction~(FSI) a natural place to look: it can modify phases after
	the short-distance production step and thereby turn a moderate
	\SUF-breaking source into a larger channel-dependent effect.  A useful
	precedent is provided by two-body hadronic $D$-meson decays, where
	long-distance rescattering and exchange topologies are known to be important
	and can generate sizable effective \SUF\ breaking
	\cite{Cheng:2010ry,Cheng:2012xb,Geng:2024uxp}.
	We therefore ask whether a common baryon--antibaryon FSI, combined with the 
	\SUF-breaking production, can account for the data. The left panel of
	Fig.~\ref{fig:production-fsi-schematic} shows the corresponding
	short-distance three-gluon production topology, while the right panel
	illustrates, for instance, the rescattering of $N\bar N$ to
	$\Sigma\bar\Sigma$. 

	\begin{table}[t]
		\centering
		\caption{Experimental inputs for $\psi\to\bb\bbar$ with
			$\psi=\jpsi,\psip$.  The $\jpsi$ branching fractions are in
			units of $10^{-3}$, while the $\psip$ branching fractions are in
			units of $10^{-4}$~\cite{ParticleDataGroup:2024cfk}.  The phase $\Delta\Phi$ is quoted in degrees.
			}
		\label{tab:experimental-inputs}
		\begin{tabular}{|c|c|c|c|c|}
			\hline
			Channel
			&
			State
			&
			${\cal B}$
			&
			$\alpha$
			&
			$\Delta\Phi$
			\\
			\hline
				\multirow{2}{*}{$\Sigma^0\bar\Sigma^0$}
				& $\jpsi$\cite{BESIII:2017kqw,BESIII:Sigma0PRL2024}
				& $1.172(32)$
				& $-0.413(9)$
				& $-4.74(43)$
				\\
				& $\psip$\cite{BESIII:2017kqw,BESIII:Sigma0PRL2024}
				& $2.35(9)$
				& $0.814(40)$
				& $29.3(5.2)$
				\\
				\hline
				\multirow{2}{*}{$\Sigma^+\bar\Sigma^-$}
				& $\jpsi$\cite{BESIII:SigmaPlus2025,BESIII:SigmaPlusBranching}
				& $1.07(4)$
				& $-0.505(2)$
				& $-15.72(20)$
				\\
				& $\psip$\cite{BESIII:SigmaPlus2025,BESIII:SigmaPlusBranching}
				& $2.43(10)$
				& $0.713(11)$
				& $24.47(1.27)$
			\\
			\hline
			\multirow{2}{*}{$p\bar p$}
			& $\jpsi$\cite{BESIII:JpsiNNbar}
			& $2.121(29)$
			& $0.595(19)$
			& -- 
			\\
			& $\psip$\cite{BESIII:psi3686NNbar}
			& $2.94(8)$
			& $1.03(7)$
			& --
			\\
			\hline
			\multirow{2}{*}{$n\bar n$}
			& $\jpsi$\cite{BESIII:JpsiNNbar}
			& $2.09(16)$
			& $0.50(21)$
			& --
			\\
			& $\psip$\cite{BESIII:psi3686NNbar}
			& $3.06(15)$
			& $0.68(16)$
			& --
			\\
			\hline
				\multirow{2}{*}{$\Lambda\bar\Lambda$}
				& $\jpsi$\cite{BESIII:2017kqw,BESIII:LambdaPRL2022}
				& $1.89(9)$
				& $0.475(4)$
				& $43.09(45)$
				\\
				& $\psip$\cite{BESIII:2017kqw,BESIII:LambdaPsi2S2025}
				& $3.81(13)$
				& $0.830(22)$
				& $21.0(3.8)$
				\\
				\hline
				\multirow{2}{*}{$\Xi^0\bar\Xi^0$}
				& $\jpsi$\cite{BESIII:Xi0Pairs,BESIII:Xi0JpsiPhase2023}
				& $1.17(4)$
				& $0.514(16)$
				& $66.9(1.5)$
				\\
				& $\psip$\cite{BESIII:Xi0Pairs,BESIII:Xi0Psi2SPhase2026}
				& $2.3(4)$
				& $0.768(38)$
				& $14.7(3.5)$
				\\
				\hline
				\multirow{2}{*}{$\Xi^-\bar\Xi^+$}
				& $\jpsi$\cite{BESIII:XiChargedPairs,BESIII:XiMinusJpsi2026}
				& $0.98(4)$
				& $0.585(6)$
				& $69.9(1.0)$
				\\
				& $\psip$\cite{BESIII:XiChargedPairs,BESIII:XiMinusPsi3686Polarization}
				& $2.87(11)$
				& $0.693(69)$
				& $38.2(7.2)$
			\\
			\hline
		\end{tabular}
	\end{table}

	Branching fractions and angular
	distributions for $\jpsi$ and $\psip$ decays to octet baryons are now
	measured with good precision, and BESIII polarization analyses have made
	several electric--magnetic relative phases accessible.
	The helicity-amplitude and timelike form-factor descriptions are
	standard~\cite{Jacob:1959at,Cabibbo:1961sz,Denig:2012by,Pacetti:2014jai,Faldt:2017kgy}.
	Existing theoretical treatments include perturbative hard scattering, mass
	corrections, \SUF\ and electromagnetic amplitudes, factorization in quantum
	chromodynamics, and
	explicit rescattering~\cite{Claudson:1981fj,Carimalo:1985mw,Murgia:1994dh,LopezCastro:1994xw,
	Bolz:1997as,Chen:2006yn,Zhu:2015bha,BaldiniFerroli:2019abd,
	Ferroli:2020mra,Mo:2021asa,Kivel:2022fzk}.  The measured relative phases provide
	information beyond rates and make it possible to separate flavor structure
	from final-state phases.

	The negative value of $\alpha_\Sigma^{\jpsi}$ indicates a sizable
	channel-dependent change in the relative ${\rm S}$- and ${\rm D}$-wave
	content.  We test this possibility in a minimal framework in which the two
	charmonia have independent real production amplitudes but share a common
	baryon--antibaryon FSI operator.  The setup follows the Watson-type
	separation of a short-distance production kernel from final-state
	phases~\cite{Watson:1952ji,Omnes:1958hv}.
	The remainder of this paper is organized as follows.  In
	Sec.~\ref{sec:formalism}, we define the helicity amplitudes, the flavor
	basis, and the common ${\rm S}$--${\rm D}$ FSI parametrization.  In
	Sec.~\ref{sec:results}, we present the isospin-reduced inputs and the reduced
	fits.  We conclude in Sec.~\ref{sec:conclusion} with implications for the
	negative $\Sigma\bar\Sigma$ angular parameter and future measurements.

	\section{Formalism}
	\label{sec:formalism}
	
	Consider the decays $\psi\to\bb\bbar$.  The field $\bb$ denotes the standard matrix representation of the light-baryon \SUF\ octet 
	\begin{equation}
	\bb^i_{j}
	=
	\begin{pmatrix}
		\frac{\Sigma^0}{\sqrt2}+\frac{\Lambda}{\sqrt6}
		&
		\Sigma^+
		&
		p
		\\[0.4em]
		\Sigma^-
		&
		-\frac{\Sigma^0}{\sqrt2}+\frac{\Lambda}{\sqrt6}
		&
		n
		\\[0.4em]
		-\Xi^-
		&
		\Xi^0
		&
		-\frac{2\Lambda}{\sqrt6}
	\end{pmatrix}. 
	\label{eq:baryon-octet}
	\end{equation}
	The indices $i,j=1,2,3$ label the $u,d,s$ flavor components.
	For each partial wave $L={\rm S},{\rm D}$, we use one \SUF-singlet
	contraction and two structures generated by the octet spurion
	$\overline m={\rm diag}(1,1,-2)$: 
	\begin{align}
		|{\bf 1}\rangle
		&=
		\frac{1}{\sqrt{8}}
		|\bbar^i_j \bb^j_i\rangle ,
		\\
		|{\bf 8}_a\rangle
		&=
		\frac{\overline m^j_k}{6}
		\left(
		|\bbar^i_j \bb^k_i\rangle
		-
		|\bbar^k_i \bb^i_j\rangle
		\right),
		\\
		|{\bf 8}_s\rangle
		&=
		\frac{\overline m^j_k}{\sqrt{20}}
		\left(
		|\bbar^i_j \bb^k_i\rangle
		+
		|\bbar^k_i \bb^i_j\rangle
		\right),
	\end{align}
	The factors $6$ and $\sqrt{20}$ normalize the octet states.  In the \SUF\ limit, a
	flavor-singlet $\psi$ produces only the
	${\bf 1}$ component.  The ${\bf 8}_a$ and ${\bf 8}_s$ components
	therefore represent explicit \SUF-breaking productions induced by the
	spurion $\overline m$. 
		For a given charmonium state $\psi$ and partial wave $L$, the direct-transition
	amplitudes are collected as
	\begin{equation}
		\vec{\cal A}^{\,\psi}_{L,{\rm dir}}
=
\left(
\begin{array}{c}
			\calA_{L,{\rm dir}}^{\psi,{\bf 1}} 
	\\[.2em]
			\calA_{L,{\rm dir}}^{\psi,{\bf 8}_a}
	\\[.2em]
			\calA_{L,{\rm dir}}^{\psi,{\bf 8}_s}
\end{array}
\right) 
		=
			\calA_{L,{\rm dir}}^{\psi,{\bf 1}} 
		\left(
		\begin{array}{c}
1 
			\\[.2em]
		r_L^{\psi,{\bf 8}_a}
			\\[.2em]
		r_L^{\psi,{\bf 8}_s}
		\end{array}
		\right). 
	\end{equation}
		These reduced production amplitudes are taken to be real because they
		correspond to short-distance transitions.  Observable complex phases are
		generated by the common FSI.  For each charmonium state, we collect the
		S- and D-wave amplitudes as 
		\begin{equation}
			\vec{\cal V}_{{\rm dir}}^{\,\psi}
			=
			\left(
\begin{array}{c} 
			\vec{\cal A}^{\,\psi}_{{\rm S},{\rm dir}}\\
					\vec{\cal A}^{\,\psi}_{{\rm D},{\rm dir}}
\end{array}
			\right),
		\end{equation}
		and use the all-order transformation
			\begin{equation}
				\vec{\cal V}_{\rm phys}^{\,\psi}
				=
				\exp(i\Phi_{\rm FSI})\,
				\vec{\cal V}_{{\rm dir}}^{\,\psi},
			\qquad
				\Phi_{\rm FSI}
				=
					\begin{pmatrix}
					\Phi_{{\rm S},{\rm FSI}} & {\cal K}_{{\rm SD}}\\
					{\cal K}_{{\rm SD}} & \Phi_{{\rm D},{\rm FSI}}
					\end{pmatrix}.
					\label{eq:full-fsi-exponential}
				\end{equation}
		For each partial wave $L$, the flavor block of the dimensionless FSI
		phase matrix is written in the same $({\bf 1},{\bf 8}_a,{\bf 8}_s)$
		basis as
	\begin{equation}
		\Phi_{L,{\rm FSI}}
		=
		\left(
		\begin{array}{ccc}
				\delta_{L0} & c_{La} & c_{Ls}
			\\[.25em]
				c_{La} & \delta_{La} & 0
			\\[.25em]
				c_{Ls} & 0 & \delta_{Ls}
		\end{array}
		\right)_{({\bf 1},{\bf 8}_a,{\bf 8}_s)}.
	\end{equation}
		The diagonal entries $\delta_{L0},\delta_{La},\delta_{Ls}$ are elastic
		phases, while $c_{La}$ and $c_{Ls}$ describe \SUF-breaking
		singlet--octet mixings.  We take these
			mixing entries to be real in the leading closed-channel
			approximation motivated by one-meson exchange.  Complex off-diagonal entries would
		parametrize higher-order corrections.  The direct FSI
	between ${\bf 8}_a$ and ${\bf 8}_s$ is ignored because it requires two
		\SUF-breaking insertions and is therefore second order in the
		breaking expansion.  We count the direct-transition ratios
		$r_L^{\psi,{\bf 8}_i}, c_{La}/\pi, c_{Ls}/\pi=O(\epsilon)$ 
		as first-order FSI breaking insertions.  The diagonal phases
		$\delta_{L0},\delta_{La},\delta_{Ls}$ are \SUF-conserving
		entries and may be larger than the singlet--octet mixings entries.
	If the FSI is closed under $\bb\bbar\to\bb'\bbar'$ rescattering
	\cite{Haidenbauer:2014kja}, $\Phi_{\rm FSI}$ is Hermitian and
$ 
	(\vec{\cal V}^{\,\psi}_{{\rm phys}})^\dagger
	\vec{\cal V}^{\,\psi}_{{\rm phys}}
	=
	(\vec{\cal V}^{\,\psi}_{{\rm dir}})^\dagger
	\vec{\cal V}^{\,\psi}_{{\rm dir}} .
$  
	This norm conservation is not identical to conservation of the
	phase-space-weighted sum of the measured branching fractions.
\begingroup
					In the reduced numerical benchmark, we take the ${\rm S}$--${\rm D}$
		mixing block to be
\begin{equation}
								{\cal K}_{{\rm SD}}
						=
				{\rm diag}
				\left(
				\KSD^{\bf 1},
				\KSD^{{\bf 8}_a},
				\KSD^{{\bf 8}_s}
					\right).
\end{equation}
		This keeps the diagonal ${\rm S}$--${\rm D}$ rescattering entries
		needed in the numerical benchmark below, while still neglecting the
		off-diagonal ${\bf 8}_a$--${\bf 8}_s$ FSI entries, which are
		second order in the breaking expansion.
\endgroup

	For a physical $\bb\bbar$ channel, we define the flavor projection
	coefficients
	\begin{equation}
		g_{\bf B}^{\bf 1}
		=
		\langle \bb\bbar|{\bf 1}\rangle,
		\qquad
		g_{\bf B}^a
		=
		\langle \bb\bbar|{\bf 8}_a\rangle,
		\qquad
		g_{\bf B}^s
		=
		\langle \bb\bbar|{\bf 8}_s\rangle.
	\end{equation}
		These Clebsch-type factors project reduced amplitudes onto physical charge
		or isospin-reduced channels and are derived from
		Eq.~\eqref{eq:baryon-octet}.  Explicitly,
\begin{equation}
	\begin{pmatrix}
		g_N^{\bf 1} & g_\Lambda^{\bf 1} & g_\Sigma^{\bf 1} & g_\Xi^{\bf 1}
		\\[6pt]
		g_N^{a} & g_\Lambda^{a} & g_\Sigma^{a} & g_\Xi^{a}
		\\[6pt]
		g_N^{s} & g_\Lambda^{s} & g_\Sigma^{s} & g_\Xi^{s}
	\end{pmatrix}
	=
	\begin{pmatrix}
		\dfrac{1}{\sqrt{8}} & \dfrac{1}{\sqrt{8}} & \dfrac{1}{\sqrt{8}} & \dfrac{1}{\sqrt{8}}
		\\[6pt]
		\dfrac{1}{2} & 0 & 0 & -\dfrac{1}{2}
		\\[6pt]
		-\dfrac{1}{\sqrt{20}} & -\dfrac{1}{\sqrt{5}} & \dfrac{1}{\sqrt{5}} & -\dfrac{1}{\sqrt{20}}
	\end{pmatrix}.
	\label{eq:flavor-projection-coefficients-matrix} 
\end{equation} 
				The corresponding projected ${\rm S}$-wave and reduced ${\rm D}$-wave
				amplitudes used in the fit are
	\begin{align}
		S_{\bf B}^\psi
		&=
		g_{\bf B}^{\bf 1}\calA_{S,\rm phys}^{\psi,{\bf 1}}
		+
		g_{\bf B}^a\calA_{S,\rm phys}^{\psi,{\bf 8}_a}
		+
		g_{\bf B}^s\calA_{S,\rm phys}^{\psi,{\bf 8}_s},
		\\
		D_{\bf B}^\psi
		&=
		g_{\bf B}^{\bf 1}\calA_{D,\rm phys}^{\psi,{\bf 1}}
		+
		g_{\bf B}^a\calA_{D,\rm phys}^{\psi,{\bf 8}_a}
		+
		g_{\bf B}^s\calA_{D,\rm phys}^{\psi,{\bf 8}_s}. 
	\end{align}

		With the standard form-factor normalization, the partial width is
		\begin{equation}
			\Gamma_{\bf B}^\psi
			=
			\frac{|\vec p_{\bf B}^{\,\psi}|}{6\pi}
			\left(
			|G_{M,{\bf B}}^\psi|^2
			+
			\frac{2m_{\bf B}^2}{m_\psi^2}|G_{E,{\bf B}}^\psi|^2
			\right), 
	\end{equation}
	where
	\begin{equation} 
		|\vec p_{\bf B}^{\,\psi}|
		=
		\frac{1}{2}
		\sqrt{m_\psi^2-4m_{\bf B}^2}.
		\label{eq:momentum-two-body}
\end{equation} 
To convert widths to branching fractions, we use
		$\Gamma_{\jpsi}=92.6\,{\rm keV}$ and
		$\Gamma_{\psip}=293\,{\rm keV}$~\cite{ParticleDataGroup:2024cfk}.
		Writing
		$P=p_{\bf B}+p_{\overline{\bf B}}$,
		$Q=p_{\bf B}-p_{\overline{\bf B}}$, we use~\cite{Faldt:2017kgy} 
		\begin{equation}
				{\cal M}
			=
			\bar u(p_{\bf B}) 
			\left[
			G_{M,{\bf B}}^\psi\gamma_\mu
			-
			\frac{2m_{\bf B}}{Q^2}
			\left(G_{M,{\bf B}}^\psi-G_{E,{\bf B}}^\psi\right)Q_\mu
			\right]
			v 
			(p_{\overline{\bf B}} )
			\,\epsilon^\mu_\psi\,,
			\label{eq:faldt-vertex}
		\end{equation}
			where the form factors are evaluated at $P^2=m_\psi^2$.
			The partial wave amplitudes are related to the form factors by~\cite{Jacob:1959at,Haidenbauer:2014kja}:
				\begin{equation}
					S_{\bf B}^\psi
					=
					\left(
					G_{M,{\bf B}}^\psi
					+
					\frac{m_{\bf B}}{m_\psi}G_{E,{\bf B}}^\psi
					\right),
					\qquad
					D_{\bf B}^\psi
					=
					\frac{m_\psi^2}{\sqrt{2}|\vec p_{\bf B}^{\,\psi}|^2}
					\left(
					G_{M,{\bf B}}^\psi
					-
					\frac{2m_{\bf B}}{m_\psi}G_{E,{\bf B}}^\psi
					\right).
				\end{equation} 
	The angular parameter $\alpha_{\bf B}^\psi$ is defined by~\cite{Jacob:1959at,Faldt:2017kgy,Du:2024jfc},
	\begin{equation}
		\alpha_{\bf B}^\psi
		=
		\frac{
			m_\psi^2 |G_{M,{\bf B}}^\psi|^2
			 - 4m_{\bf B}^2  |G_{E,{\bf B}}^\psi|^2
		}{
		m_\psi^2 	|G_{M,{\bf B}}^\psi|^2
			+ 4m_{\bf B}^2 |G_{E,{\bf B}}^\psi|^2
		}.
	\end{equation}
	The relative phase measured through polarization observables is
	\begin{equation}
		\Delta\Phi_{\bf B}^\psi
		=
		\arg\left(
		\frac{G_{E,{\bf B}}^\psi}{G_{M,{\bf B}}^\psi}
		\right). 
	\end{equation}
		Thus 
		$\alpha_{\bf B}^\psi$ fixes the relative magnitude of $G_E$ and $G_M$, and
		$\Delta\Phi_{\bf B}^\psi$ their phase.

\section{Numerical results and discussions}
\label{sec:results}

	After charge partners are combined, we work with the $\mathrm{SU}(2)_F$-exact
	isospin channels $N,\Lambda,\Sigma,\Xi$.  The input then contains 22
	isospin-reduced data points: 16 values of ${\cal B}$ and $\alpha$ plus
	six measured relative phases.  These inputs are collected in
	Table~\ref{tab:prediction-comparison}, together with the fit values and
	predictions used below.

\begin{table}[!htbp]
\centering
\caption{Isospin-reduced data and predictions for the ${\bf 8}_a$-active fit with full diagonal $\KSD$.  Branching-fraction
units follow Table~\ref{tab:experimental-inputs}, and $\Delta\Phi$ is in
degrees.}
\label{tab:prediction-comparison} 
\begin{tabular}{|c|c|c c|c c|c c|}
\hline
\multicolumn{2}{|c|}{} & \multicolumn{2}{c|}{${\cal B}$} &
\multicolumn{2}{c|}{$\alpha$} &
\multicolumn{2}{c|}{$\Delta\Phi$} \\
\hline
${\bf B}$ & $\psi$ &
Data & Fit &
Data & Fit &
Data & Fit \\
\hline
\multirow{2}{*}{$N$}
& $\jpsi$\cite{BESIII:JpsiNNbar}
& $2.12(3)$ & $2.12(4)$
& $0.59(2)$ & $0.59(3)$
& -- & $-15(6)$ \\
& $\psip$\cite{BESIII:psi3686NNbar}
& $2.97(7)$ & $3.0(1)$
& $0.97(6)$ & $0.87(6)$
& -- & $16(13)$ \\
\hline
\multirow{2}{*}{$\Lambda$}
& $\jpsi$\cite{BESIII:2017kqw,BESIII:LambdaPRL2022}
& $1.89(9)$ & $1.9(1)$
& $0.475(4)$ & $0.475(6)$
& $43.1(5)$ & $43.1(6)$ \\
& $\psip$\cite{BESIII:2017kqw,BESIII:LambdaPsi2S2025}
& $3.81(13)$ & $3.8(1)$
& $0.83(2)$ & $0.85(3)$
& $21(4)$ & $22(4)$ \\
\hline
\multirow{2}{*}{$\Sigma$}
& $\jpsi$\cite{BESIII:2017kqw,BESIII:Sigma0PRL2024, BESIII:SigmaPlus2025,BESIII:SigmaPlusBranching}
& $1.13(3)$ & $1.13(4)$
& $-0.500(2)$ & $-0.500(3)$
& $-13.8(2)$ &$-13.8(3)$ \\
& $\psip$\cite{BESIII:2017kqw,BESIII:Sigma0PRL2024, BESIII:SigmaPlus2025,BESIII:SigmaPlusBranching}
& $2.39(7)$ & $2.4(1)$
& $0.72(1)$ & $0.72(2)$
& $24.7(1.2)$ & $25(2)$ \\
\hline
\multirow{2}{*}{$\Xi$}
& $\jpsi$\cite{BESIII:Xi0Pairs,BESIII:Xi0JpsiPhase2023,BESIII:XiChargedPairs,BESIII:XiMinusJpsi2026}
& $1.08(3)$ & $1.07(4)$
& $0.578(5)$ & $0.578(8)$
& $69.1(8)$ & $69(1)$ \\
& $\psip$\cite{BESIII:Xi0Pairs,BESIII:Xi0Psi2SPhase2026,BESIII:XiChargedPairs,BESIII:XiMinusPsi3686Polarization}
& $2.83(11)$ & $2.8(1)$
& $0.75(3)$ & $0.72(4)$
& $19(3)$ & $19(5)$ \\
\hline
\end{tabular}
\end{table}

\begingroup
	The parameter count is organized around the \SUF-exact singlet limit.  This
	baseline has the 4 real singlet production amplitudes
	$\calA_{{\rm S},{\rm dir}}^{\psi,{\bf 1}}$ and
	$\calA_{{\rm D},{\rm dir}}^{\psi,{\bf 1}}$, with $\psi=\jpsi,\psip$,
	together with 3 diagonal ${\rm S}$--${\rm D}$ rescattering parameters
	$\KSD^{\bf 1}$, $\KSD^{{\bf 8}_a}$, and $\KSD^{{\bf 8}_s}$.
	First-order \SUF\ breaking in production adds the 2 octet
	structures ${\bf 8}_a$ and ${\bf 8}_s$ for each $\psi$ and each
	partial wave, giving 8 additional real direct-transition amplitudes.  The
	common FSI block contains 6 diagonal phases, while first-order \SUF-breaking
	FSI adds 4 singlet--octet mixing entries.  After fixing the unphysical
	overall phase by $\delta_{{\rm S}0}=0$, a fully unrestricted common-FSI fit
	therefore contains $4+8+6+4+3-1=24$ real parameters.
	This parameter count motivates reduced common-FSI scenarios.  In addition to
	the \SUF-exact case, we consider an
	${\bf 8}_a$-active case with
		$\calA_{L,{\rm dir}}^{\psi,{\bf 8}_s}=0$.  This assumption removes 4
		direct-production parameters.  With the diagonal rescattering block used
		above, the reduced fit has 20 parameters and 2 nominal degrees of freedom.
\endgroup
	
Numerically, the \SUF-exact scenario gives
$\chi^2\simeq 10^5$, showing that \SUF\ breaking is needed to explain the
data.  To keep the reduced amplitudes natural, we impose $|r|\leq0.49$ in
the $\chi^2$ fit.  The ${\bf 8}_a$-active scenario with full diagonal
$\KSD$ gives $\chi^2=4.21$~\footnote{The corresponding
${\bf 8}_s$-active full-diagonal fit gives $\chi^2=58.79$.}.
The parameter values in this scenario are collected in
Table~\ref{tab:fitted-parameters}.  The sizable fitted
$\KSD^{{\bf 8}_s}$ shows that the symmetric octet still participates through
the common rescattering block rather than through direct production.
The fit results in Table~\ref{tab:prediction-comparison} show that
$\alpha_{\Sigma}^{\jpsi}$ is reproduced within this framework.

\begin{table}[H]
\centering
\caption{Best-fit parameters for the ${\bf 8}_a$-active fit with full
diagonal $\KSD$.  Phases are in radians,   and parentheses give local
one-standard-deviation errors.}
\label{tab:fitted-parameters}
\setlength{\tabcolsep}{7pt}
\renewcommand{\arraystretch}{1.12}
\begin{tabular}{|l|c|l|c|}
\hline
\multicolumn{4}{|c|}{FSI parameters } \\
\hline
$\delta_{{\rm S}a}$ & $0.93(21)$&
$\delta_{{\rm S}s}$ &$2.82(6)$ \\
$\delta_{{\rm D}0}$ & $-0.21(18)$ &
$\delta_{{\rm D}a}$ & $-3.05(11)$ \\
$\delta_{{\rm D}s}$ &$-0.12(5)$ &
 & \\
$c_{{\rm S}a}$ & $-0.18(5)$ &
$c_{{\rm S}s}$ & $-0.57(7)$ \\
$c_{{\rm D}a}$ & $-0.82(27)$ &
$c_{{\rm D}s}$ & $-0.41(2)$ \\
$\KSD^{\bf 1}$ & $0.10(1)$ &
$\KSD^{{\bf 8}_a}$ &  $-0.30(2)$\\
$\KSD^{{\bf 8}_s}$ &  $-1.3(1)$ &
& \\
\hline
\multicolumn{4}{|c|}{Direct transition} \\
\hline
$\calA_{{\rm S},{\rm dir}}^{\jpsi,{\bf 1}}$ & $-1.3(2)\times10^{-3}$ &
$r_{\rm S}^{\jpsi,{\bf 8}_a}$ &$0.34(35)$ \\
$\calA_{{\rm D},{\rm dir}}^{\jpsi,{\bf 1}}$ & $2.25(14)\times10^{-2}$ &
$r_{\rm D}^{\jpsi,{\bf 8}_a}$ & $0.32(21)$ \\
$\calA_{{\rm S},{\rm dir}}^{\psip,{\bf 1}}$ & $-3.48(8)\times10^{-3}$ &
$r_{\rm S}^{\psip,{\bf 8}_a}$ & $-0.12(4)$ \\
$\calA_{{\rm D},{\rm dir}}^{\psip,{\bf 1}}$ & $-8.7(7)\times10^{-3}$ &
$r_{\rm D}^{\psip,{\bf 8}_a}$ & $-0.41(21)$ \\
\hline
\end{tabular}
\end{table}

To display the mechanism more explicitly, we expand the projected
${\rm S}$-wave and reduced ${\rm D}$-wave amplitudes used in the fit.
Using the central values in Table~\ref{tab:fitted-parameters}, the
separate ${\rm S}$- and ${\rm D}$-wave flavor-block exponentials are
\begingroup
\begin{align}
\exp(i\Phi_{{\rm S},{\rm FSI}})
&\simeq
\begin{pmatrix}
0.908-0.103\,i & 0.071-0.149\,i & 0.367-0.058\,i\\
0.071-0.149\,i & 0.585+0.793\,i & -0.014-0.039\,i\\
0.367-0.058\,i & -0.014-0.039\,i & -0.906+0.198\,i
\end{pmatrix},
\label{eq:s-fsi-exponential}
\\[0.4em]
\exp(i\Phi_{{\rm D},{\rm FSI}})
&\simeq
\begin{pmatrix}
0.790+0.027\,i & -0.484+0.042\,i & -0.037-0.371\,i\\
-0.484+0.042\,i & -0.857+0.116\,i & -0.058+0.110\,i\\
-0.037-0.371\,i & -0.058+0.110\,i & 0.913-0.110\,i
\end{pmatrix}.
\label{eq:d-fsi-exponential}
\end{align}
\endgroup
In the exact \SUF\ limit the off-diagonal entries of
$\Phi_{L,{\rm FSI}}$ would disappear.  The nonzero off-diagonal entries in
Eqs.~\eqref{eq:s-fsi-exponential} and~\eqref{eq:d-fsi-exponential} are therefore
the fitted FSI source of flavor mixing, and each remains below the 50\% level in
magnitude.

The ${\rm D}$-wave block
has sizable off-diagonal components; after the common $\KSD$ block is
included, this allows ${\rm D}$-wave strong phases to feed the
${\rm S}$-wave amplitudes.
The $\jpsi$ amplitudes read 
\begin{equation}
\begin{aligned}
S_{N}^{\jpsi}
\simeq{}& -4.7\times10^{-4}\Big[
1+\underbrace{(0.5)}_{\Delta_r}
+\underbrace{(0.2-4.8\,i)}_{\Delta_{\rm FSI}}
+\underbrace{(0.5+i\,1.0)}_{\Delta_{\rm rest}}
\Big],
\\
D_{N}^{\jpsi}
\simeq{}& 8.0\times10^{-3}\Big[
1+\underbrace{(0.4)}_{\Delta_r}
+\underbrace{(-0.8+i\,0.3)}_{\Delta_{\rm FSI}}
+\underbrace{(-1.0+i\,0.06)}_{\Delta_{\rm rest}}
\Big],
\\
S_{\Sigma}^{\jpsi}
\simeq{}& -4.7\times10^{-4}\Big[
1+\underbrace{(0.0)}_{\Delta_r}
+\underbrace{(2.6-i\,0.04)}_{\Delta_{\rm FSI}}
+\underbrace{(-0.06-i\,0.23)}_{\Delta_{\rm rest}}
\Big],
\\
D_{\Sigma}^{\jpsi}
\simeq{}& 8.0\times10^{-3}\Big[
1+\underbrace{(0.0)}_{\Delta_r}
+\underbrace{(-0.3-i\,0.4)}_{\Delta_{\rm FSI}}
+\underbrace{(-0.2+i\,0.06)}_{\Delta_{\rm rest}}
\Big].
\end{aligned}
\label{eq:jpsi-sd-decomposition}
\end{equation}
The corresponding $\psip$ amplitudes are
\begin{equation}
\begin{aligned}
S_{N}^{\psip}
\simeq{}& -1.2\times10^{-3}\Big[
1+\underbrace{(-0.2)}_{\Delta_r}
+\underbrace{(-0.2+i\,0.4)}_{\Delta_{\rm FSI}}
+\underbrace{(0.2+i\,0.01)}_{\Delta_{\rm rest}}
\Big],
\\
D_{N}^{\psip}
\simeq{}& -3.1\times10^{-3}\Big[
1+\underbrace{(-0.6)}_{\Delta_r}
+\underbrace{(-0.8+i\,0.4)}_{\Delta_{\rm FSI}}
+\underbrace{(1.3-i\,0.07)}_{\Delta_{\rm rest}}
\Big],
\\
S_{\Sigma}^{\psip}
\simeq{}& -1.2\times10^{-3}\Big[
1+\underbrace{(0.0)}_{\Delta_r}
+\underbrace{(-0.10-i\,0.16)}_{\Delta_{\rm FSI}}
+\underbrace{(-0.02-i\,0.01)}_{\Delta_{\rm rest}}
\Big],
\\
D_{\Sigma}^{\psip}
\simeq{}& -3.1\times10^{-3}\Big[
1+\underbrace{(0.0)}_{\Delta_r}
+\underbrace{(-0.4-i\,0.4)}_{\Delta_{\rm FSI}}
+\underbrace{(0.2-i\,0.08)}_{\Delta_{\rm rest}}
\Big].
\end{aligned}
\label{eq:psip-sd-decomposition}
\end{equation}
Here $\Delta_r$ is the direct ${\bf 8}_a$ correction with FSI switched
off, $\Delta_{\rm FSI}$ is the correction from applying the common FSI to the
singlet source alone, and $\Delta_{\rm rest}$ is the remaining interference
between direct ${\bf 8}_a$ production and FSI.  Comparing $N$ with
$\Sigma$ makes the role of the antisymmetric octet transparent:
$\Delta_r$ is visible in the $N\bar N$ amplitudes, while it vanishes for
$\Sigma\bar\Sigma$ because $g_\Sigma^a=0$.  The negative
$\alpha_\Sigma^{\jpsi}$ is therefore generated by the common FSI acting on
the ${\rm S}$- and ${\rm D}$-wave amplitudes, not by a direct
$\Sigma$-channel ${\bf 8}_a$ source.  The fitted amplitudes in
Table~\ref{tab:fitted-parameters} are ${\rm D}$-wave dominated, most
clearly for $\jpsi$, where the singlet reduced ${\rm D}$-wave baseline is
about an order of magnitude larger than the projected ${\rm S}$-wave
baseline. 

This provides a simple answer to the apparent large
\SUF-breaking question.  The entry $\Delta_{\rm FSI}\simeq2.6$ in
$S_\Sigma^{\jpsi}$ is not an independent large
$\Sigma$-channel flavor-breaking amplitude; it can be decomposed into three
effects.  First, the fitted ${\rm D}$-wave source is much larger than the
${\rm S}$-wave source.  Second, the \SUF-breaking correction remains modest
when it is normalized within the ${\rm D}$-wave amplitude itself.  Third, the
common $\KSD$ block transfers this ${\rm D}$-wave correction into
$S_\Sigma^{\jpsi}$ through ${\rm D}\to{\rm S}$ rescattering.  Since the
$S_\Sigma$ baseline is small, the transferred correction becomes large only
after normalization to the ${\rm S}$-wave bracket.  The apparent large \SUF\
breaking is therefore a normalized-amplitude enhancement, not evidence for a
large $\Sigma$-specific \SUF-breaking source.
When these amplitudes are converted back to $G_E$ and $G_M$, they also
	reproduce the phase clue in Table~\ref{tab:prediction-comparison}:
	$\Delta\Phi_{\Sigma}^{\jpsi}\simeq -13.8^\circ$, while the corresponding
	$\jpsi$ phases for $\Lambda\bar\Lambda$ and $\Xi\bar\Xi$ are positive.
Thus the same rescattered ${\rm S}$- and ${\rm D}$-wave pattern that drives
the negative $\alpha_{\Sigma}^{\jpsi}$ also makes
$\Sigma\bar\Sigma$ carry a different strong phase from the other measured
$\jpsi$ channels.

The largest remaining tension in the fit is the
$\psip\to N\bar N$ angular observable: $\alpha_N^{\psip}$ contributes
$\Delta\chi^2\simeq2.48$.  Since the isospin-reduced $N\bar N$ input
averages the charge modes in Table~\ref{tab:experimental-inputs}, this pull
should be read as a charge-resolved issue.  The table gives
$\alpha_{p}^{\psip}=1.03(7)$ and $\alpha_{n}^{\psip}=0.68(16)$, a roughly
$2.0\sigma$ internal tension.  Within the present framework, the
$\psip\to n\bar n$ angular distribution is therefore the most direct place
to revisit experimentally, because it is tied to the largest deviation left by
the fit.  By comparison, the $\Sigma\bar\Sigma$ rows in
Table~\ref{tab:prediction-comparison} are fitted with subleading pulls while
retaining the negative $\jpsi$ angular parameter. The fitted $N\bar N$ phases in Table~\ref{tab:prediction-comparison}
are not used as measurement priorities, since they are not directly measurable
in the present $N\bar N$ channels. Charge-resolved polarization measurements, together with a revisit of
$\psip\to n\bar n$, can further test the same mechanism at BESIII and at the
future Super Tau-Charm Facility~\cite{Achasov:2023gey}.

Before closing, we stress that the fit never requires a first-order
\SUF-breaking ingredient larger than the 50\% level when expressed in the
quantities used to display the mechanism.  In the FSI matrices of
Eqs.~\eqref{eq:s-fsi-exponential} and~\eqref{eq:d-fsi-exponential}, all
off-diagonal flavor-mixing entries are below 50\% in magnitude, and the direct
production ratios in Table~\ref{tab:fitted-parameters} obey $|r|<0.5$.  The
large-looking entry in $S_\Sigma^{\jpsi}$ is therefore not a large direct
\SUF-breaking ratio; it is the normalized result of the common FSI transferring
the modest ${\rm D}$-wave breaking into a small ${\rm S}$-wave baseline.

\section{Conclusion}
\label{sec:conclusion}

We have studied the negative-$\Sigma\bar\Sigma$ puzzle in
$\jpsi,\psip\to\bb\bbar$ angular distributions.  The data show
$\alpha_\Sigma^{\jpsi}<0$, whereas the other measured $\jpsi$ octet
channels and the corresponding $\psip$ channels have positive angular
parameters.  In a singlet plus octet basis with a common FSI, the
\SUF-exact singlet description is strongly disfavored, giving
$\chi^2\simeq10^5$.

Allowing the common FSI to mix the ${\rm S}$- and
${\rm D}$-wave amplitudes, with an antisymmetric direct octet and a full
diagonal $\KSD$ block, gives $\chi^2=4.21$ for two nominal degrees of
freedom.  The fit values and predictions are shown in
Tables~\ref{tab:prediction-comparison} and~\ref{tab:fitted-parameters}.
The negative $\Sigma\bar\Sigma$ angular parameter is traced to the
different rescattered ${\rm S}$- and ${\rm D}$-wave pattern in
$\jpsi\to\Sigma\bar\Sigma$.  The direct ${\bf 8}_a$ source affects
$N\bar N$ already before rescattering, while the $\Sigma\bar\Sigma$
effect is generated through the common FSI.  The opposite sign of the measured
$\jpsi\to\Sigma\bar\Sigma$ phase relative to the measured $\Lambda$ and
$\Xi$ phases gives the same message: the puzzle is tied to strong phases.
Seen this way, the fit does not
require a large $\Sigma$-specific \SUF-breaking source.  The apparent large
breaking in $S_\Sigma^{\jpsi}$ is decomposed into three points:
${\rm D}\gg{\rm S}$, the breaking remains modest when normalized in the
${\rm D}$-wave amplitude, and ${\rm D}\to{\rm S}$ rescattering through the
common $\KSD$ block moves this correction into the small $S_\Sigma$
baseline.  It therefore appears large only in the normalized
${\rm S}$-wave bracket.  This produces $\alpha_\Sigma^{\jpsi}<0$, while
the $\psip$ production amplitudes keep $\alpha_\Sigma^{\psip}$ positive.

The largest remaining deviation is associated with
$\alpha_N^{\psip}$, where the isospin-reduced input averages proton and
neutron angular measurements that are in mild tension.  We therefore recommend
an experimental revisit of $\psip\to n\bar n$, especially its angular
distribution, as the most direct check of the leading residual in the present
framework.  Future charge-resolved polarization observables can test the same
mechanism.

\acknowledgments

We thank Xiongfei Wang and Hongfei Shen for valuable discussions.
This work is supported in part by the National Natural Science Foundation of China (NSFC) under Grant Nos. 12547104 and
12575096, and by the China Postdoctoral Science Foundation under Grant No. 2025M773361.

\end{document}